\documentclass[12pt]{amsart}

\usepackage{amsmath,amsfonts,amsthm,amscd,amssymb,graphicx,mathrsfs}
\usepackage{cite}
\usepackage{amsbsy}
\usepackage{graphicx}
\usepackage{subcaption}
\usepackage{bm}
\usepackage{tikz}
\usetikzlibrary{arrows}
\usepackage[text={6.5in,9in},centering,includefoot,foot=0.6in]{geometry}

\usepackage[all]{xy}

\numberwithin{equation}{section}
\numberwithin{figure}{section}

\usepackage{setspace}
\definecolor{skyblue}{rgb}{0.85,0.85,1}

\newtheorem{theorem}{Theorem}[section]
\newtheorem{lemma}[theorem]{Lemma}

\newtheorem{conjecture}[theorem]{Conjecture}

\theoremstyle{definition}

\newtheorem{define}[theorem]{Definition}

\theoremstyle{remark}
\newtheorem{remark}[theorem]{Remark}

\newcommand{\bC}{\mathbb{C}}

\newcommand{\R}{\mathbb{R}}
\newcommand{\T}{\mathbb{T}}
\newcommand{\Z}{\mathbb{Z}}

\newcommand{\cB}{\mathcal{B}}

\newcommand{\cF}{\mathcal{F}}
\newcommand{\cH}{\mathcal{H}}

\newcommand{\Ga}{\Gamma}

\newcommand{\be}{\begin{equation}}
\newcommand{\ee}{\end{equation}}
\newcommand{\bel}{\begin{lemma}}
\newcommand{\el}{\end{lemma}}
\newcommand{\bt}{\begin{theorem}}
\newcommand{\et}{\end{theorem}}
\newcommand{\bec}{\begin{conjecture}}
\newcommand{\ec}{\end{conjecture}}
\newcommand{\bed}{\begin{define}}
\newcommand{\ed}{\end{define}}
\newcommand{\ber}{\begin{remark}}
\newcommand{\er}{\end{remark}}

\begin{document}

\title[Dispersion relations]{Analytic and algebraic properties of dispersion relations (Bloch varieties) and Fermi surfaces.\\What is known and unknown}

\author{Peter Kuchment}
\address{Department of
  Mathematics, Texas A\&M University, College Station, TX 77843-3368, USA}
\email{kuchment@tamu.edu}
%\homepage{http://www.math.tamu.edu/~kuchment}
%\address{Mathematics Department, Texas A\&M University, College Station, TX
%77843-3368, USA}

\thanks{P.K. acknowledges support of NSF DMS-2007408 grant.}
\date{\today}
\begin{abstract}
The article surveys the known results and conjectures about the analytic properties of dispersion relations and Fermi surfaces for periodic equations of mathematical physics and their spectral incarnations.
\end{abstract}

\maketitle
%\parskip=0em{\tableofcontents}
%\tableofcontents

%\parskip=1ex
\section{Introduction}
This brief survey mostly follows and extends the outlines given in the author's lectures at the conferences ``Learning from Insulators
New Trends in the Study of Conductivity of Metals'' at the Lorenz
Center in Leiden, Netherlands (August 2021) and ``Algebraic Geometry in Spectral Theory'' at ICERM, Brown University (February 2023).

Dispersion relations and Fermi surfaces for periodic operators of mathematical physics are some of the most common and important notions in condensed matter physics, in particular in the lately immensely popular nano-materials (graphene, nanotubes, etc.), photonic crystal theory, and topological insulators. We attempt here to survey their basic analytic properties, their relations to the spectra of the corresponding operators, their physical meaning, and some related results and conjectures.

Since the readership of this text might come from diverse communities of mathematicians and physicists, the author apologizes in advance for providing some discussions that might seem obvious to some, but not necessarily to others. So, for a mathematician, who has not dealt with dispersion relations, we start with analogy of constant coefficients partial differential equations/operators (PDEs). We then switch to honest periodic operators and introduce the corresponding definitions. After that we start discussing the analytic properties of these varieties and their role for properties of solutions and the spectrum.

The word ``algebraic'' in the title is justified by the fact that in the discrete situations (e.g., the popular tight binding approximation) the arising problems of analytic geometry become the ones of algebraic geometry instead. This is certainly an ``improvement,'' since one can use powerful results of algebraic geometry, sometimes not available in the analytic case, On the other hand, the discrete (graph) case sometimes presents surprises impossible in the continuous case, due to non-trivial graph topology and absence of the uniqueness of continuation property. A significant progress in the discrete case has been occurring lately.

Regretfully, it would be impossible to provide any proofs here. So, the reader is referred to the literature for those.
Many references can be found in the survey \cite{KuchBAMS16}. However, since a lot has happened in the past several years, we'll try to present many more citations.
As always, even with the best of intentions, some important references might be missed, and the author offers his apologies in advance.

%%%%%%%%%%%%%%%%%%%%%%%%
\section{Constant coefficient PDEs and Fourier transform}
%%%%%%%%%%%%%%%%%%%%%%
We denote by  $D$ the operator
\be
D:=\frac1i\nabla
\ee
in $\R^d$, where $\nabla=\dfrac{\partial}{\partial x}$ is the standard gradient. Any linear constant coefficient operator $L$ of order $m$ can be written as a polynomial in $D$:
\be
L(D)=\sum\limits_{|\alpha|\leq m}D^\alpha, \, L=L(D):u(x)\mapsto L(D)u(x).
\ee
Here $\alpha$ is a multi-index consisting of the orders of differentiation with respect to each coordinate.

After Fourier transform $u(x)\mapsto \hat{u}(\xi)$, one has
\be L: \hat{u}(\xi)\mapsto L(\xi)\hat{u}(\xi),
\ee
where the polynomial
\be
L(\xi)=\sum\limits_{|\alpha|\leq m}\xi^\alpha
\ee
is called the \textbf{symbol} of $L$. This is essentially the simplest case of the \textbf{direct integral decomposition}
\be
L=\int\limits^\oplus_{\xi\in\R^d}L(\xi).
\ee

%%%%%%%%%%%%%%%%%
\subsection{``Dispersion relation'' and ``Fermi surface'' for constant coefficient operators}
%%%%%%%%%%%%%%%%
The quotation marks here reflect the fact that no one uses these words in the constant coefficient case. Well, we are in the free world and will use them.
\bed\indent
\begin{itemize}
\item The graph of $L(\xi)$ (in $\R^d\times\R$ or $\bC^d\times\bC$)
\be
\cB:=\{(\xi,\lambda)\,|\, \lambda=L(\xi)\},
\ee
i.e. the graph of the symbol, is \textbf{``dispersion relation''}, or \textbf{``Bloch variety''}.
\item The level surface
\be
\cF_\lambda:=\{\xi\,|\, L(\xi)=\lambda\}
\ee
is \textbf{``Fermi surface''} at the ``energy level'' $\lambda$.
    \end{itemize}
\ed
Note that using not only real, but also complex values of $\xi$ and $\lambda$ turns out to be important, both in the constant coefficient and periodic situations.

The following statement is clear, because both of these varieties are given as sets of zeros of polynomials:
\bel
Both Bloch and Fermi varieties are algebraic.
\el

Notice that the ``Fermi'' surface $\cF_\lambda$ is just the set of zeros of the symbol $L(\xi)-\lambda)$ It is called the \textbf{characteristic set} for the operator $L(D)-\lambda$ and is known to carry many ``secrets'' about the operator (see, e.g. \cite{HormPDE}).

%%%%%%%%%%%%%%%%%%%%%
\subsection{Systems of constant coefficient PDEs}
%%%%%%%%%%%%%%%%%%%%%%%%%%%
Let $L(D)$ be a system, i.e. a matrix of linear constant coefficient differential operators. Then the symbol $L(\xi)$ is also a matrix. Thus, the direct integral expansion becomes
\be
L=\int\limits^\oplus_{\R^d}L(\xi).
\ee
Things seem to be very much parallel to the scalar case, except that it might be cumbersome (albeit sometimes useful) to use the ``graph'' of the matrix-valued symbol $L(\xi)$. Thus, \textbf{losing some information}, one graphs just its spectrum $\sigma(L(\xi))$. Now we deal with a multiple-valued scalar function, and thus we arrive to the following definition:
\bed\indent
\begin{itemize}
\item \textbf{``Dispersion relation''}, or \textbf{``Bloch variety''} (in $\R^d\times\R$ or $\bC^d\times\bC$ is
$$\cB:=\{(\xi,\lambda)\,|\, \det (L(\xi)-\lambda I)=0\},$$
i.e., $\cB$ is the graph of the multiple-valued function $\xi\mapsto\sigma(L(\xi))$
\item The \textbf{``Fermi surface''} at the energy level $\lambda$:
$$\cF_\lambda:=\{\xi\,|\, (\xi,\lambda)\in \cB\}$$
\end{itemize}
\ed
Algebraicity of both varieties still holds, since one is looking for the set of zeros of the polynomial
\be
D(\xi,\lambda):=\det \left(L(\xi)-\lambda I\right).
\ee
And now we are closer to introducing the corresponding definitions for the periodic operators.

%%%%%%%%%%%%%%%%%%
\section{Periodicity enters the fry}
%%%%%%%%%%%%%%%%%

%%%%%%%%%%%%%
\subsection{Group of periods}
%%%%%%%%%%%%%%%%%%%
Our starting point in the constant coefficient case was using the Fourier transform. This was natural, since such operators commute with the group $\Gamma=\R^d$ acting (transitively) on $\R^d$ by shifts. Then $L(\xi)$ is the restriction of $L$ to the $1D$ space generated by the function $e^{i\xi\cdot x}$, which is how any irreducible representation (\textbf{irrep}) of $\Ga$ looks like. This is just the common spectral direct integral decomposition into irreps (see, e.g., \cite{Dixmier}).

Let us see now what happens in the periodic case. Let $\Ga=\Z^d$ act (non-transitively) on $\R^d$ by shifts\footnote{The specific choice of a co-compact abelian sub-group of shifts is irrelevant at this, and many others, moment.}. Irreps of $\Ga$ are $e^{ik\cdot \gamma}$, with \textbf{quasi-(crystal- )momentum}
\be
k\in \R^d (\mbox{ or }\bC^d) \, mod\, \Gamma^*,
\ee
where $\Gamma^*:=2\pi\Z^d$ is the \textbf{dual lattice} to $\Ga=\Z^d$.
The corresponding eigenspace $\cH_k$ for $\Ga$ consists of \textbf{Bloch functions} of the form
\be
v(x)=e^{ik\cdot x}p(x),
\ee
where $p$ is $\Gamma$-periodic.

We fix choices of a \textbf{fundamental domain} $W$ (\textbf{Wigner-Seitz cell}) for action of $\Ga$ on $\R^d$ and of a fundamental domain $B$ (\textbf{Brillouin zone}) for action of $\Ga^*$. Functions that are $\Ga$-periodic can be considered as functions on the torus $\T:=\R^d_x/\Ga$. Analogously, $\Ga^*$-periodic functions can be identified with functions on the ``Brillouin torus''  $\T^*:=\R^d_k/\Ga^*$. These tori are supplied with the natural \textbf{flat} metrics pushed down from the space $\R^2$. We will denote the corresponding normalized to the volume $1$ measures by $dx$ and $dk$ correspondingly.

All $\Ga$-invariant linear partial differential operators are of the form
\be
L=L(x,D)=\sum a_\alpha (x)D^\alpha
\ee
with $\Ga$-periodic functions $a_\alpha$.

If we denote by $L(k)$ the operator $L$ acting on the functions from $\cH_k$, then we have the following direct integral expansion
\be\label{E:directint}
L=\int\limits^\oplus_{k\in T^*}L(k)dk
\ee
A convenient representation of $L(k)$ for a periodic operator $L=L(x,D)$ is the following:
%%%%%%%
\bel
$L(k)=L(x,D+k)$ acting on function on the torus $T:=\R^d/\Ga$.
\el
%%%%%
In particular, if $L=-\Delta+V(x)=(D)^2 +V(x)$, then
$$L(k)=(\frac1i\nabla+k)^2+V(x)=-\Delta+\frac{2}{i}k\cdot\nabla+k^2+V(x),$$
acting on $\Ga$-periodic functions (i.e., functions on $\T$).

\ber\indent
\begin{itemize}
\item The advantage of the operators $L(k)$ in comparison with $L$ is that they act on a compact manifold (torus) and if they are elliptic, their spectra are discrete.
\item Since the correspondence between quasimomenta $k$ and irreps of $\Ga$ is not one-to-one, it is often convenient to introduce the    \textbf{Floquet multiplier}
\be
z:=e^{ik}:=(e^{ik_1},\dots,e^{ik_d})\in(\bC\setminus \{0\})^d.
\ee
If $k$ is real, $z$ in ``Brillouin torus'' $\T^*:=\R^d/\Ga^*$.

One can observe that in the discrete case, in the Floquet multipliers rather than quasimomenta representation, the Bloch and Fermi varieties become algebraic.
\end{itemize}
\er

\ber
Before going further, the reader should be warned that the irrep expansions in the (often important) case of co-compact action of a \textbf{discrete non-abelian (=non-commutative) group}\footnote{Needed, for instance, to allow for presence of constant magnetic field. The corresponding magnetic translations group of \textbf{Zak transformations} is non-abelian (see, e.g. the \cite{Auslander_nilmanifolds,Zak}).} $\Ga$ does not seem too useful, due to complicated structure of the space of the irreducible representations, and in particular due to usage of non-unitary representations (corresponding to the complex quasi-momenta $k$) in the abelian case. It has been understood that the better language is the one of group algebras (not providing any complete reference, we just mention as token examples \cite{GruberNoncom,GrunBlochQ,WannierThiang}).

We will not address the non-abelian case, which is a different animal, here.
\er

%%%%%%%%%%%%%%%
\section{The class of operators}
%%%%%%%%%%%%%%%%%%

Let us describe precisely the classes of operators under consideration. Here are main assumptions (although sometimes we will need to weaken or strengthen them):
\indent\begin{itemize}
\item \be\label{E:operator}
L=L(x,D)=\sum\limits_{|\alpha|\leq m}a_\alpha (x) D^\alpha,
\ee
where the coefficients $a_\alpha$ are $\Ga$-periodic and smooth\footnote{Infinite differentiability of the coefficients is a significant overkill, but we will not address the issues of minimal conditions on the coefficients.}
\item The operator is assumed to be \textbf{elliptic}, i.e. its principal symbol
\be
L_0(x,\xi):= \sum\limits_{|\alpha|= m}a_\alpha (x) \xi^\alpha
\ee
does not vanish for any real non-zero vectors $\xi$.

This is a crucial assumption, without which the tools and results of the whole analytic theory of Bloch and Fermi varieties essentially fall apart or have to be changed  significantly \footnote{One can handle to some extent the hypo-elliptic, e.g. parabolic equations, while things get much harder there \cite{Kuc_floquet}.}. This applies for instance for time-periodic wave or Schr\"odinger type equations (see for some discussions, applications, and references \cite{Kuc_floquet,KuchBAMS16,WeinstTimeFloquet,FefWeinsDirac,YajimaSurv,SviridovComplet,Zeld,Gavrila} and references therein).
\item It will be often assumed that the operator is self-adjoint in $L_2(\R^d)$, although some basic facts, like for instance analyticity of the varieties, hold without this condition.
\item In some results one will need to require that the operator is of the second order $m=2$. Indeed, some of the crucial statements do not hold for general periodic elliptic operators of higher order, due to possible absence of unique continuation property \cite{Plis,Kuc_floquet,KuchBAMS16}. The same issue is encountered in the periodic discrete or quantum graph cases \cite{KuchBAMS16}.
  
\item Some other (matrix) operators of mathematical physics, such as Maxwell or Dirac ones are also important and studied (see e.g. \cite{Morame_acMaxwell,BirSusAveragMax,BirSusDirAC2,Danilov_00,Danilov_05,Danilov_11,KucLev_tams02,Kuch_pbg}), but we will not address these in the current text.
\end{itemize}
 Our {\bf canonical example} here is the Schr\"odinger operator with real periodic electric potential
 \be\label{E:schr}
 L=-\Delta +V(x).
 \ee

%%%%%%%%%%%%%%%%%%%%%%%
\section{And finally, Bloch and Fermi varieties!}
%%%%%%%%%%%%%%%%%%
\bed
The \textbf{Bloch variety (dispersion relation)} of operator $L$
$$\cB_L:=\{(k,\lambda)|\,\lambda\in\sigma(L(k)) \}.$$
The \textbf{Fermi surface} at the energy level $\lambda$
$$\cF_\lambda:=\{k|\,\lambda\in\sigma(L(k))\}.$$
\ed
\begin{figure}[ht!]
\begin{center}
  \includegraphics[scale=0.7]{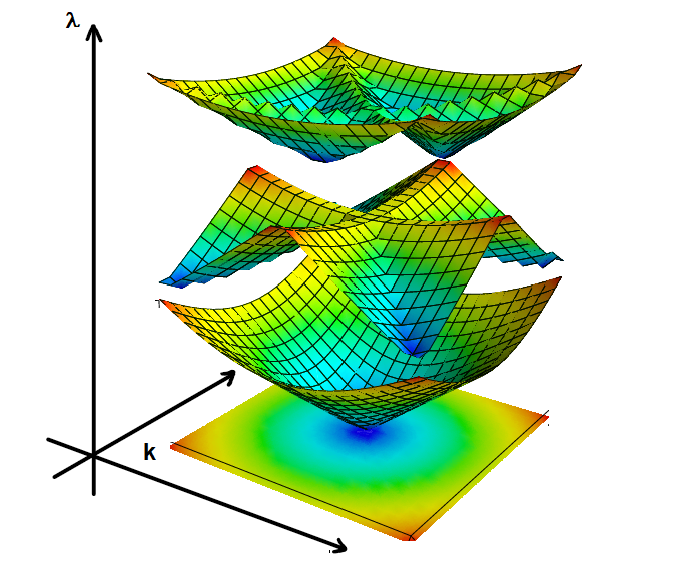}\includegraphics[scale=0.4]{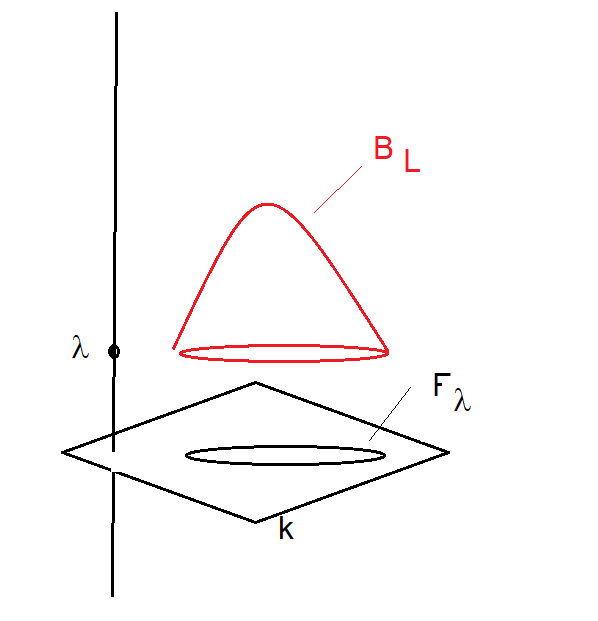}\\
  \caption{Cartoons of Bloch variety (left) and Fermi surface (right).}\label{F:cartoons}
\end{center}
\end{figure}
\begin{center}
\begin{figure}[ht!]
  \includegraphics[scale=0.7]{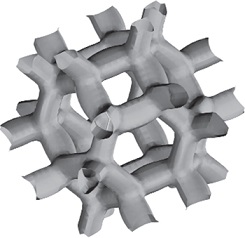}  \includegraphics[scale=0.4]{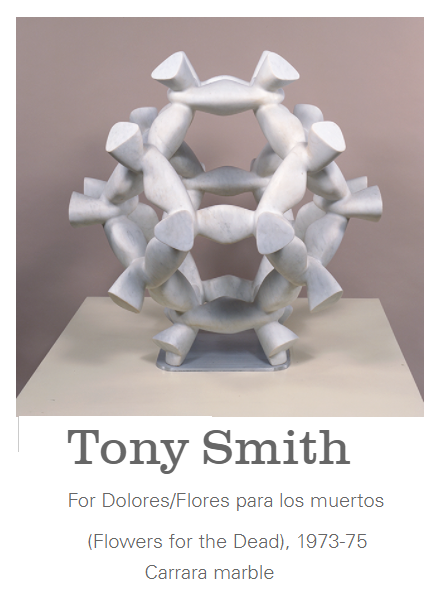}
\caption{Beautiful Fermi surface of Pb}\label{F:Pb}
\end{figure}
\end{center}
\ber
The reader should note that the above definitions of the Bloch and Fermi varieties do not specify whether the quasimomenta $k$ and eigenvalues $\lambda$ have to be real. One can \textbf{allow them to be complex }without a problem, to obtain the \textbf{complex versions} $\cB_\bC \subset \bC^{d+1}, \cF_\lambda \subset \bC^2$.
\er

%%%%%%%%%%%%%%%%%%%%
\section{Analyticity of Bloch and Fermi varieties}
%%%%%%%%%%%%%%%%%%
We have seen in the case of constant coefficient equations and systems of those that the Bloch and Fermi varieties were ``nice'', namely algebraic sets (i.e., were the sets of zeros of the polynomial with respect to $(k,\lambda)$ function $\det (L(\xi)-\lambda I)$).
It would have be nice if we could have a similar description in the periodic case, something like ``$\det (L(k)-\lambda I)=0$.'' However, our operators $L(k)$ are unbounded for each $k$. Here is where ellipticity gets into the game. The operators $L(k)$ now have discrete spectra, but a simple-minded attempt to define some ``determinants'' for them fails. One can use, however, the so called \textbf{regularized determinants} (see \cite{Kuc_floquet,GohbergKrein_nonselfadjoint,Dunford}, which save the day and enable one to prove the following result \cite{Kuc_floquet,KuchBAMS16}:
%%%%%%%%%%%%%%%%%
\bt
There exists an entire function $f(k,\lambda)$ on $\bC^{d+1}$, such that
\be
\cB=\{(k,\lambda)\in\ \bC^{d+1}|\, f(k,\lambda)=0\},
\ee
\be
\cF_\lambda=\{k\in\ \bC^{d}|\, f(k,\lambda)=0\},
\ee
\be\label{E:estim}
|f(k,\lambda)|\leq Ce^{C(|k|+|\lambda|)^r}
\ee
for some $C>0$ and $r\geq 1$.
\et

Our analogy with the constant coefficient case is almost achieved, except that the varieties of interest happen to be analytic, rather than algebraic sets. In other words, they can be described as the sets of zeros of entire functions, rather than polynomials. In some questions this does not make that much of a difference, while some useful properties of algebraic sets just do not survive for the analytic ones. In particular, the following ``almost all or almost nothing'' property of algebraic sets comes handy: projection of an algebraic set onto a either is contained in a proper algebraic subset (i.e., is ``almost nothing''), or contains the complement to such a subset (i.e., is almost ``everything'').
Regretfully, the zero sets of entire functions might not have this property. Fortunately, having estimates of the sort of (\ref{E:estim}) helps \cite{Lelong,Eremenko} and raises some hopes.

%%%%%%%%%%%%
\section{The spectrum}
%%%%%%%%%%%%%%

If the operator (\ref{E:operator}) is bounded from below and self-adjoint, then the eigenvalues of $L(k)$ for real $k$ can be labeled in non-decreasing order\footnote{This labeling disappears when $k$ is complex.}
\be\label{E:bandfunct}
\lambda_1(k)\leq \lambda_2(k)\leq \dots \to \infty.
\ee
The \textbf{band functions} $\lambda_j(k)$ are continuous and piece-wise analytic, losing analyticity only when they collide.

Due to the direct decomposition (\ref{E:directint}) one has
\be\label{E:projspect}
\sigma(L)=\bigcup\limits_{k\in B}\sigma (L(k)).
\ee
The following theorem summarizes well known results about the spectra of periodic elliptic operators (see, e.g., \cite[Vol. 4]{ReedSimon} and \cite{Kuc_floquet,Eastham_periodic,KuchBAMS16,Skr_psim85}).%This proves that
\bt \label{T:spectrumproj}\indent\begin{enumerate}
\item The spectrum $\sigma(L)$ is the projection on the spectral (i.e., $\lambda$-) axis of the real Bloch variety $\cB_L$.
\item The spectrum consists of the \textbf{spectral bands} $I_j$, which are the ranges of the band functions $\lambda_j(k)$ and thus are closed finite intervals.
\item When $j\to\infty$, both endpoints of $I_j$ tend to infinity.
\item In dimension higher than $1$, the spectral bands might overlap.
\item The spectral bands might leave some uncovered \textbf{spectral gaps} in between. The spectral gaps ends (edges) correspond to extrema of the dispersion relation $\lambda(k)$.
\end{enumerate}
\et

%%%%%%%%%%%%
\section{Bloch and Fermi varieties and spectrum}
%%%%%%%%%%%%%%%%%
\subsection{Geometry of the spectrum - bands and gaps}
According to the statement (1) of Theorem \ref{T:spectrumproj}, the Bloch variety (dispersion relation) determines the spectrum and tells us that it has the band-gap structure. A very important questions is whether the spectral gaps do exist. Fort instance, existence of semi-conductors and photonic crystals depends upon this property \cite{AshcroftMermin_solid,Kuch_pbg,Joannopoulos_photocrystals}.

The spectrum is usually purely absolutely continuous. At least, this expected to be true for all second order elliptic periodic operator. Absence of singular continuous spectrum is often a rather mundane conclusion of periodicity \cite{KuchBAMS16}, and is in fact much more general - it holds for a class of analytically fibered operators \cite{GerNie_jfa98}, which our direct integrals provide.  A different story is presence (or rather absence) of the pure point spectrum, i.e. of bound states.
\begin{conjecture}
The spectrum of any periodic elliptic second order operator with sufficiently ``nice'' coefficients is absolutely continuous, and thus no bound states exist.
\end{conjecture}
This result has been proven for SChr\:odinger operators \ref{E:schr} almost half of a century ago in \cite{Tho_cmp73} (see also expositions in \cite{ReedSimon,Kuc_floquet,KuchBAMS16}). Proving this in presence of a periodic magnetic potential turned out to be much harder, took a long time, and was finally achieved in \cite{Sob_inv99}. Handling variable coefficients in the second order terms would not succumb to the previously used harmonic analysis tools. This has been achieved in \cite{Fri_cmp02}, where a seemingly irrelevant symmetry conditions was imposed, but no one has succeeded to remove it.

Why the second order restriction? The reason is that there are examples of periodic elliptic operators of higher order (of elasticity type), which do have bound states. The reason is the absence of week uniqueness of continuation property, and thus existence of compactly supported solutions (see discussion in \cite{KuchBAMS16}). 
The role of the (weak) uniqueness continuation principle\footnote{It is saying that a solution vanishing in an open set is identically zero.} is emphasized by the 2nd order operator example provided in \cite{Filonov_example}, in which the coefficients in the leading term are just a notch below those guarantying uniqueness of continuation, and indeed, compactly supported solutions (and thus bound states) do appear.
Graphs (discrete and quantum) notoriously lack uniqueness of continuation, and thus bound states can exist. Moreover, in this case if a bound state does exist, then the corresponding eigenspace is generated by compactly supported solutions \cite{Kuc_incol89,KuchBAMS16,BerKuc_book,Kui_t65,KuchZhao}. If such a result could be proven in the continuous case, it would have proven absolute continuity of the spectrum in one shot, due to uniqueness of continuation.\\
Another indication here is the following result \cite{Kuc_floquet,KuchBAMS16}:
\begin{theorem}
Existence of a bound state for an elliptic periodic equation implies existence of a super-exponentially decaying solution $u(x)$, i.e. such that
\be
|u(x)|\leq C e^{-C|x|^r} 
\ee
for some $r>1$ (estimates on $r$ are also available).
\end{theorem}
\begin{conjecture}
Existence of such a solution for second order periodic (or more general class) elliptic equations should be prohibited (at least in the self-adjoint case) by some kind of ``uniqueness of continuation at infinity'' (the \textbf{Landis} problem). Thus, the spectra must be absolutely continuous.
\end{conjecture}
The only results in this direction known to the author are in \cite{Mesh_decrease,FroHerHofHof_prse83}. They apply only to Schr\"doniger operators with bounded electric potentials and also allow $r$ to reach up to 4/3 (at least in the non-selfadjoint case). There is significant activity nowadays around the Landis problem, but so far none of the results produce progress in establishing the above conjecture.

An interesting and very useful observation from \cite{Tho_cmp73} (although it is not formulated there in such terms) is the following:
\begin{theorem}\label{T:flatband}
Existence of a bound state ($L_2$-eigenfunction)  at an energy $\lambda$ is equivalent to the Fermi surface $\cF_\lambda$ being ``flat'':
\be
\cF_\lambda=\bC^d_k.
\ee 
\end{theorem}
This allows one to prove absence of bound states by smartly choosing a quasimomentum value $k_0$ with large imaginary part, so that the point $k_0$ is not in $\cF_\lambda$.
%%%%%%%%%%%%%%%%
\subsection{Gaps existence and creation} 
%%%%%%%%%%%%%%%%%%%%
Fortunately, the Nature has provided us with semi-conductors, but the story of creating photonic spectral gaps was not an easy one (both experimentally and theoretically).

What could be techniques for creating spectral gaps? For a schr\"odinger operator spectral gaps can be created near the bottom of the spectrum using the following simple procedure: adding to the free operator (Laplacian) a localized deep potential well will produce an eigenvalue below the continuous spectrum. Now, repeating this well periodically at a sufficiently large distance will spread this eigenvalue into a small spectral band, thus still leaving a gap.

This procedure does not work for Maxwell operators, where there is no way to shift the bottom of the spectrum. This was exactly what the struggle for creating photonic crystals was about.

\begin{figure}{ht!}
\centering
\includegraphics[width=0.25\textwidth]{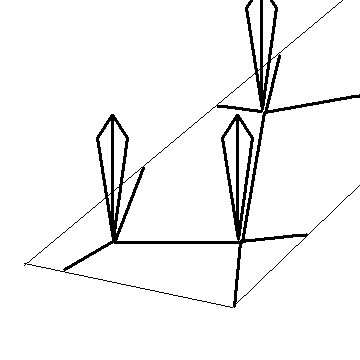}\includegraphics[width=0.25\textwidth]{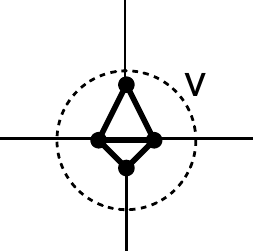}
\caption{``Kite'' (left) and ``spider'' (right) decorations.}\label{F:decorat}
\end{figure}

Another procedure is of creating periodically placed high contrast inclusions throughout the medium ( a version of the above procedure). It is known that this can create gaps throughout the spectrum (see, e.g. \cite{FigKuc_siamjam96,FigKuc_siamjam96a,FigKuc_siamjam98,SmyshKuch,Zhikov_gaps,HemLie_cpde00}).
The idea is to spread throughout the medium small identical resonators with internal structure, so near their eigenfrequencies the propagation is suppressed. Such gaps are often called \textbf{resonant}. This idea goes back at least to \cite{Pav_tmp84}, but its explicit implementations has been achieved so far only in the graph cases \cite{SchAiz_lmp00} and quantum graph \cite{BerKuc_book,Ong_diss,DoKuchOng}. The gap opening is obtained by ``decorating the graph: either by ''attaching to every graph vertex another graph (``kite'') \cite{SchAiz_lmp00} or inserting an internal structure (``spider'') into each vertex \cite{DoKuchOng}, see Fig. \ref{F:decorat}. The spider decorations look in some sense more practical, but they does not necessarily opens gaps and thus is harder to handle\footnote{It is interesting to compare the spider decoration with the zig-zag product used to produce expander graphs \cite{WidgZig_anm02}.}.

The old \textbf{Bethe-Sommerfeld conjecture} (BSC) \cite{Bethe_elektronen} claims that in dimension higher than $1$, the number of spectral gaps is expected to be finite. After a long series of works of distinguished mathematicians, this was proven for the Schr\"odinger operator case in \cite{Parn_BS}, where and in \cite{Parn_Sob_Invent} one can find further references. 

Simple counterexamples to BSC are easy to construct (by decorations) for periodic discrete graphs, although for the quantum graph case the conjecture is still open.

%%%%%%%%%%%%%%%%%
\subsection{Gap edges location}
%%%%%%%%%%%%%%%%%

For quite some time there has been a belief that the spectral gap edges must occur only at some ``symmetric'' values of quasimomenta in Brillouin zone. This assumption allows for significant dimension reduction when computing the spectrum as the set (while not resolving the density of the spectrum). This is true and has simple origin in the one-dimensional case. There is no such reason in higher dimension, and indeed, this has been disproved in \cite{ExnKucWin_jpa10,HarKucSob_jpa07}. There are, however, cases when this does hold true \cite{berCanz}.

%%%%%%%%%%%%%%%%%
\subsection{Wave packets behavior}
%%%%%%%%%%%%%%%%%%%%
It is well known that the propagation of narrow wave packets of Bloch waves, located at a point of the dispersion relation, is governed for some length of time by the constant coefficient operator whose symbol is the jet (a lower degree part of the Taylor expansion) of the dispersion relation at this point \cite{FefWeiPackets,AshcroftMermin_solid} (kind of quasi-classical limit).

There are two especially important cases. 

%%%%%%%%%%%%%%%%%%
\subsubsection{Spectral gap edge}
%%%%%%%%%%%%%%%%%%%
The first one is at the edge of a spectral gap, i.e. at an extremum of the dispersion relation $k\mapsto \lambda$.
\begin{conjecture}
Generically with respect to the parameters of the (elliptic periodic) operator, e.g. for ``almost all'' periodic potentials, the edges of the spectral gaps are isolated in the $k$-space and non-degenerate (i.e. the dispersion relation has a parabolic shape).
\end{conjecture}
This condition claims that near a spectral edge the solutions behave like in the free (albeit anisotropic) space. This allows one to define the \textbf{effective masses} \cite{AshcroftMermin_solid}. It is also important for establishing asymptotics of the Green's function near the gap edges, Liouville theorems, Anderson localization, and other issues \cite{KuchBAMS16,KuchRaich,KucPin_tams07,KucPin_jfa01,KhaKuchRaich,KhaKuchment}.

This conjecture is expected to hold true in continuous case, but there is very little progress in proving it, except for \cite{KloRal_maa00,Col_msmf91,FilKach}. In discrete case, there are known counterexamples \cite{FilKach}, as well as various positive results in some cases \cite{ParnSht2D,DoKucSott}.

%%%%%%%%%%%%%%
\subsubsection{Dirac cones}
%%%%%%%%%%%%%

Another very popular nowadays case is the appearance of (stable) Dirac cones (``diabolic points'') in the dispersion relation.

Existence of such a cone implies that (for some time) wave packets located near the cone behave like solutions of the Dirac's equation for massless fermions \cite{FefWeinsDirac}.
The famous graphene owes many of its exciting properties to existence of these cones \cite{KatsGraph}. The mandatory appearance of them in the honeycomb geometry is established in \cite{FefWeinsDirac}, see also the nice group-theoretic justification in \cite{BerkComech}.

%%%%%%%%%%%%%%%%%
\section{Am esoteric question: irreducibility}
%%%%%%%%%%%%%%%%%%
Let $Z\subset \bC^n$be an \textbf{analytic (algebraic) variety}, i.e. it can locally be represented as the set of all common zeros of finitely many analytic (polynomial) functions $f_j(z)$.

Such a set is \textbf{reducible}, if it can be represented as the union of two strictly smaller analytic (algebraic) subsets $Z_1, Z_2\subset Z$. One should imagine the case when $f(z)$ has a non-trivial factorization $f_1(z)f_2(z)$ into the product of two analytic (polynomial) functions $f_j$.

An analytic (algebraic) set is \textbf{irreducible} if it is not reducible.
Another incarnation of the irreducibility is that the set of all its smooth points is connected.

Yet another view on irreducibility is that any small open patch of the variety uniquely determines the whole of it (by analytic continuation).

The following result was established in \cite{KnoTruDircomp_cmh90}:
\begin{theorem}\label{T:irred2D}
Let $L=-\Delta +V(x)$ , where $V(x)$ is continuous periodic potential in $\R^2$. Then the Bloch variety of $L$ is irreducible.
\end{theorem}

This seems to be an esoteric statement. In fact, it is not. Indeed, it implies that
\begin{enumerate}
\item There are no bound states. I.e., the spectrum of $L$ is purely absolutely continuous.  
\item The Bloch variety shows an amazing rigidity: any small patch of a band function determines the dispersion relation. (Do not hope for doing this! It is a highly numerically unstable procedure.)
\end{enumerate}  
The proof of the first statement is simple. Indeed, as Theorem \ref{T:flatband} claims, existence of a bound states implies existence of a flat Fermi surface $\cF_\lambda$. This component shows that the Bloch variety is reducible into two component: one is $\cF_\lambda \times \{\lambda\}$ and another is the rest of the Bloch variety.

The following conjecture is believed, but still unproven (except the result mentioned above):
\begin{conjecture}
The Bloch variety of any periodic Schr\:odunger operator $-\Delta+V(x)$ (or even general second order elliptic periodic operator) is irreducible.
\end{conjecture}

In discrete case, some instances of this conjecture have been proven in \cite{Gieseker_fermi}.

The analogous question of irreducibility has been asked about the Fermi surfaces (for all, or almost all values of the energy $\lambda$). 

This irreducibility problem also seems esoteric. However, as it was discovered in \cite{KucVai_cmp06,KucVai_cpde00,KucVai_incol98}, it is at the heart of proving the absence of\textbf{ embedded states}, which is a long standing problem in dimensions 2 and higher. Namely, when a periodic medium is perturbed by a localized perturbation, the old important question is whether impurity eigenvalues can arise inside the spectral bands, rather than in the spectral gaps. This fact has been established long ago in $1D$ \cite{Rof_dansssr63,Rof_incol84,Rof_incol84} (see also \cite{,Zhel67,Zhel68,Zhel70,FirsDirInvScatt,FirsSpectIdent,FirsSpectIdent}). To the best of the author's knowledge, there had been no progress in higher dimensions, till it was discovered in \cite{KucVai_cmp06,KucVai_cpde00,KucVai_incol98} that irreducibility of the Fermi surface $\cF_\lambda$ for a given $\lambda$ inside of the continuous spectrum prevents appearance of embedded impurity eigenvalues at this point. Examples of irreducibility were presented there as well. Even before these works, conditions of Fermi surface irreducibility in the $2D$ square lattice case were found in \cite{Gieseker_fermi}. Significant recent progress proving irreducibility for discrete models, as well as examples of reducibility, and constructions of the corresponding examples of appearance of embedded eigenvalues have been provided in \cite{Papanic_emb,FillmanIrredBloch,shipman_qgr,ReducFermi,ShipmanReducible,LiuWembedded,LiuWirredFermi,LiuWtopics}.

\section{Acknowledgments}The author acknowledges the support from the NSF DMS Grant \# 2007408. Thanks also go to many colleagues for their fruitful discussion and collaboration. This includes, but not limited to G.~Berkolaiko, Ngoc Do, J.~Fillman, L.~Friedlander, W.~Liu, B.~Ong, R.~Matos, S.~Shipman, F.~Sottile, and J. Zhao.

\bibliographystyle{plain}
\bibliography{pk_periodic}

\end{document}